\documentclass[sigconf]{acmart}

\AtBeginDocument{%
  \providecommand\BibTeX{{%
    \normalfont B\kern-0.5em{\scshape i\kern-0.25em b}\kern-0.8em\TeX}}}

\usepackage{multirow}
\usepackage{subcaption}
\usepackage[title]{appendix}

\copyrightyear{2022} 
\acmYear{2022} 
\setcopyright{acmcopyright}\acmConference[UIST '22]{The 35th Annual ACM Symposium on User Interface Software and Technology}{October 29-November 2, 2022}{Bend, OR, USA}
\acmBooktitle{The 35th Annual ACM Symposium on User Interface Software and Technology (UIST '22), October 29-November 2, 2022, Bend, OR, USA}
\acmPrice{15.00}
\acmDOI{10.1145/3526113.3545674}
\acmISBN{978-1-4503-9320-1/22/10}

\usepackage{xcolor}

\newcommand\mydots{\hbox to 1em{.\hss.\hss.}}

\newcommand\remove[1]{}

\usepackage[utf8]{inputenc}

\usepackage{tabularx}
\usepackage[percent]{overpic}

\usepackage{nameref}

\usepackage{hyperref}
\usepackage{url}

\usepackage{xspace}
\usepackage{enumitem}
\usepackage{mathtools}
\usepackage{commath}

\usepackage{pifont}

\newcommand{\customtilde}{{\raise.17ex\hbox{$\scriptstyle\sim$}}}

\newcommand{\mv}[1]{\mathbf{#1}}

\newcommand{\N}{\mbox{\rm \hbox{I\kern-.15em\hbox{N}}}}
\newcommand{\R}{\mbox{\rm \hbox{I\kern-.15em\hbox{R}}}}

\def \N {\mbox{\rm \hbox{I\kern-.15em\hbox{N}}}}
\def \R {\mbox{\rm \hbox{I\kern-.15em\hbox{R}}}}

\newcommand{\bd}{\mathbf{d}}

\newcommand{\bff}{\mathbf{f}}

\newcommand{\bx}{\mathbf{x}}

\newcommand{\bq}{\mathbf{q}}

\newcommand{\energydensity}{W}
\newcommand{\materialdensity}{d}

\begin{document}
\begin{teaserfigure}
    \centering
    \includegraphics[width=\textwidth]{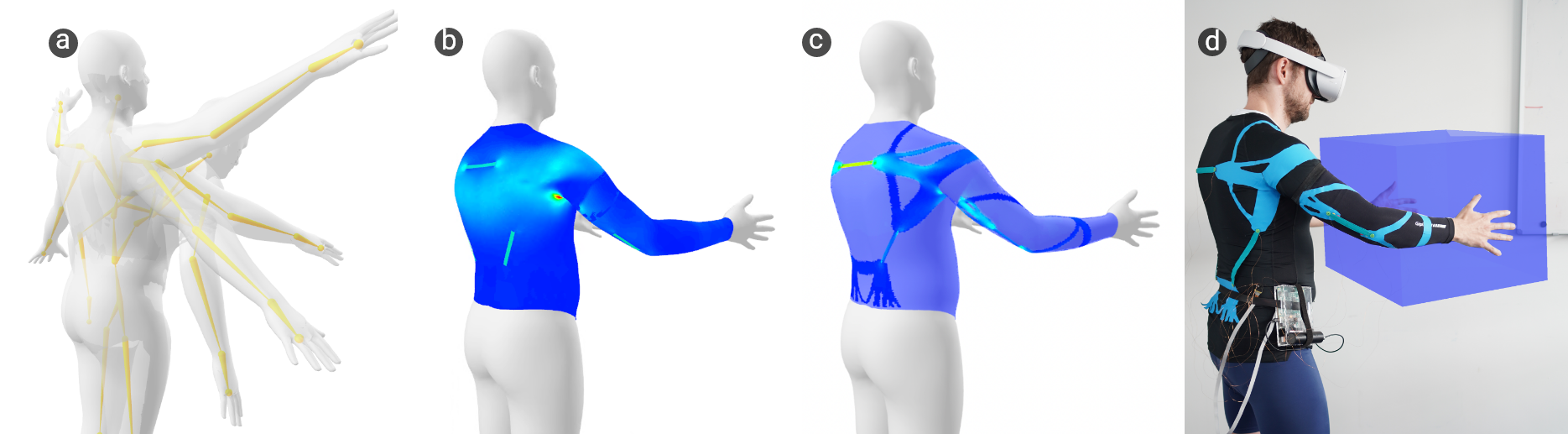}
    \caption{Our pipeline for designing active kinesthetic garments. Given a set of motions  (a) and a compliant garment with electrostatic clutches (light green) (b), our method automatically generates efficient connecting structures (c), that offer strong resistance to input motions when clutches are \textit{active} while minimizing interference otherwise. Once fabricated, our designs can provide force-feedback on-demand in various applications such as a VR training scenarios (d). }
    \label{fig:teaser}
\end{teaserfigure}

\title{Computational Design of Active Kinesthetic Garments}



\author{Velko Vechev}
\affiliation{%
  \institution{Department of Computer Science}
  \city{ETH Z{\"u}rich}
  \country{Switzerland}}
\email{velko.vechev@inf.ethz.ch}

\author{Ronan Hinchet}
\affiliation{%
  \institution{Department of Computer Science}
  \city{ETH Z{\"u}rich}
  \country{Switzerland}}
\email{ronan.hinchet@inf.ethz.ch}

\author{Stelian Coros}
\affiliation{%
  \institution{Department of Computer Science}
  \city{ETH Z{\"u}rich}
  \country{Switzerland}}
\email{stelian.coros@inf.ethz.ch}

\author{Bernhard Thomaszewski}
\affiliation{%
  \institution{Department of Computer Science}
  \city{ETH Z{\"u}rich}
  \country{Switzerland}}
\email{bthomasz@inf.ethz.ch}

\author{Otmar Hilliges}
\affiliation{%
  \institution{Department of Computer Science}
  \city{ETH Z{\"u}rich}
  \country{Switzerland}}
\email{otmar.hilliges@inf.ethz.ch}

\renewcommand{\shortauthors}{Vechev et al.}

\begin{abstract}
Garments with the ability to provide kinesthetic force-feedback on-demand can augment human capabilities in a non-obtrusive way, enabling numerous applications in VR haptics, motion assistance, and robotic control. However, designing such garments is a complex, and often manual task, particularly when the goal is to resist multiple motions with a single design. In this work, we propose a computational pipeline for designing  \emph{connecting structures} between active components---one of the central challenges in this context. We focus on electrostatic (ES) clutches that are compliant in their passive state while strongly resisting elongation when activated. Our method automatically computes optimized connecting structures that efficiently resist a range of pre-defined body motions on demand. We propose a novel dual-objective optimization approach to simultaneously maximize the resistance to motion when clutches are active, while minimizing resistance when inactive. We demonstrate our method on a set of problems involving different body sites and a range of motions. We further fabricate and evaluate a subset of our automatically created designs against manually created baselines using mechanical testing and in a VR pointing study. 
\end{abstract}

\ccsdesc[500]{Human-centered computing~Haptic devices}



\keywords{computational design, topology optimization, kinesthetic feedback}
\maketitle


\section{Introduction}
Kinesthetic garments are an efficient and non-obtrusive way of providing force feedback for human body motion. By augmenting stretchable fabric with strategically designed reinforcements, they offer targeted resistance to motions along specific directions \cite{vechev2022cdkg}. 
They are part of an emerging trend of soft robotic garments \cite{sanchez2021textile} that have the potential to assist human wearers in various ways such as during locomotion \cite{kim2019reducing, lee2018autonomous}, rehabilitation \cite{al2016wrist}, and increasing immersion in mixed reality \cite{gunther2019pneumact, rognon2018flyjacket, al2017frozen}. However, relying only on \textit{passive} mechanical structure for feedback prevents their use in such applications because they require \emph{active} feedback.


In this work, we propose a computational approach for designing \textit{active kinesthetic garments} that can resist user-defined motions \textit{on demand}. 
To implement such adaptive resistance, we rely on electrostatic clutches  \cite{hinchet2019highforce}, i.e., pre-fabricated components that provide extremely high stiffness contrast between their active and inactive states. 
Designing active kinesthetic garments then amounts to determining clutch placements, typically placed over high-strain areas, and finding a passive structure that connects and anchors the active components. Crucially, this layout should result in the garment providing maximal resistance when clutches are active, but minimally interfere with motion otherwise. Designing effective connecting structures requires the understanding of the interaction between stretchable garments in multiple states sliding over a deforming body in multiple poses, a very difficult and unintuitive task. 


To address this challenge, we formalize the design of active kinesthetic garments as an on-body topology optimization problem whose objective function explicitly balances the opposing goals for active and inactive states. 
By maximizing the difference in elastic energy between active and inactive states, our formulation encourages layouts in which clutches link disconnected parts of the passive structure. In this way, clutches leverage the passive structure to establish strong, load-carrying paths when active while maintaining freedom of movement otherwise.

We implement our formulation within a standard evolutionary optimization algorithm, and produce a set of active kinesthetic garment designs that each target multiple motions spanning different body sites. Our results indicate that designs produced with our approach are highly effective and outperform manually-designed alternatives by significant margins. To further substantiate this analysis, we manufacture a subset of our designs for experimental evaluation. Both mechanical testing and a VR pointing task indicate clear advantages for the designs created with our method. To summarize, we make the following contributions:

\begin{itemize}
  \item A \emph{computational design pipeline} for the automatic creation of active kinesthetic garments that includes a novel objective function that considers active components and multiple motions.
  \item  A set of \emph{fabricated active kinesthetic garments} built on compliant material integrating ES clutches as kinesthetic feedback components.
  \item A comprehensive \emph{evaluation} showing the effectiveness of our method in simulation, in a physical validation, and in a VR user study against manually-designed and visual baselines.
\end{itemize}

\section{Related Work} We summarize works in the areas of computational methods in garment modeling and augmentation, intersecting with hardware and devices capable of providing body-scale kinesthetic feedback.



\paragraph*{Body-scale Kinesthetic Haptic Feedback Systems}
Early work to provide kinesthetic feedback to the body used motors and hydraulic pistons to actuate heavy bulky haptic platforms. More recently, several wearable body kinesthetic feedback systems have been developed, mostly based on electromagnetic motors \cite{Chen2016, Shen2019} with rods \cite{Schiele2011, Barnaby2019} or cables \cite{Asbeck2014, Fang2020} transmission, and based on pneumatic actuators \cite{gunther2019pneumact, Delazio2018} which are soft and more comfortable at the detriment of a bulkier equipment (pumps, compressors, valves). An alternative way to provide body kinesthetic feedback are passive blocking mechanisms like vacuum jamming \cite{Choi2018} (still requiring pumps) and ES clutches \cite{Diller2016, hinchet2018dextres, Ramachandran2021, ramachandran2021arm}. In particular, ES clutches offer the advantages of being ultra-thin, light, and soft enabling the design of compliant kinesthetic garment designs. Such kinesthetic systems are typically manually designed to specifically fit a limb/joint and block a certain motion. In contrast, we leverage an automatic design method that models and simulates clutches, allowing us to accommodate any set of motions and body areas. 


\paragraph*{Topology Optimization}
Topology optimization is a powerful method used in engineering disciplines to most efficiently distribute a finite amount of material, typically to minimize compliance \cite{bendsoe2013topology, zehnder2021ntopo}. The graphics community has also combined compliance minimization with user guided input \cite{martinez2015structure, schumacher2016stenciling}. It has also been demonstrated on elastic materials \cite{Skouras13Computational} as well as structures undergoing large displacements \cite{bruns2001topology}. Closer to our work, topology optimization has moved into the on-body domain where it has been used for personalized cast design \cite{zhang2019customization} and casts designed for thermal comfort \cite{zhang2017thermal}. Most recently, Vechev et al. demonstrated the design of \emph{kinesthetic garments}, which are passively reinforced garments designed to resist a single motion \cite{vechev2022cdkg}. However, this work only formulates a single compliance minimization objective, and thus cannot be used in a setting that leverages active components. We extend this approach in two important ways, first by the addition of a dual objective that considers the active and inactive states of our components. A second important contribution is a formulation that enables optimization for multiple motions.


\paragraph*{Intelligent Garment Augmentation}
The intelligent design of garments is an emerging discipline with important applications for the general population. Computational design approaches to garment design have recently started to consider motion as a fundamental design quantity in so-called 4D garments \cite{liu2021knitting} that minimize friction and pressure via integrated knitting maps. In addition to minimizing friction during motion, Montes et al. also optimize for pressure distributions and body fit by employing a physically based model of skin-tight garments on the body \cite{montes2020computational}. Vechev et al. augment existing skin-tight clothing with passive reinforced materials to resist a single given motion, employing a more flexible model of the garment that allows cloth to slide and lift-off from the body \cite{vechev2022cdkg}.  Optimization of component placement has also been used in soft-robotic garments, in combining elastic cords, clutches, and dampers to reduce the force and power required by a person to generate lower body motion \cite{ortiz2017energy}. Evolutionary optimization techniques were employed by Gholami et al. for designing garments with optimally placed fabric sensors \cite{gholami2019lower}. Muthukumarana et al. integrated combinations of active shape-memory based components into garments allowing for actuation on clothing \cite{muthukumarana2021clothtiles}. In our work, we augment garments with active components that generate kinesthetic feedback and design supporting optimization objectives to create efficient structures connecting them. 




%


\begin{figure*}[!ht]
 \center
  \includegraphics[width=2.0\columnwidth]{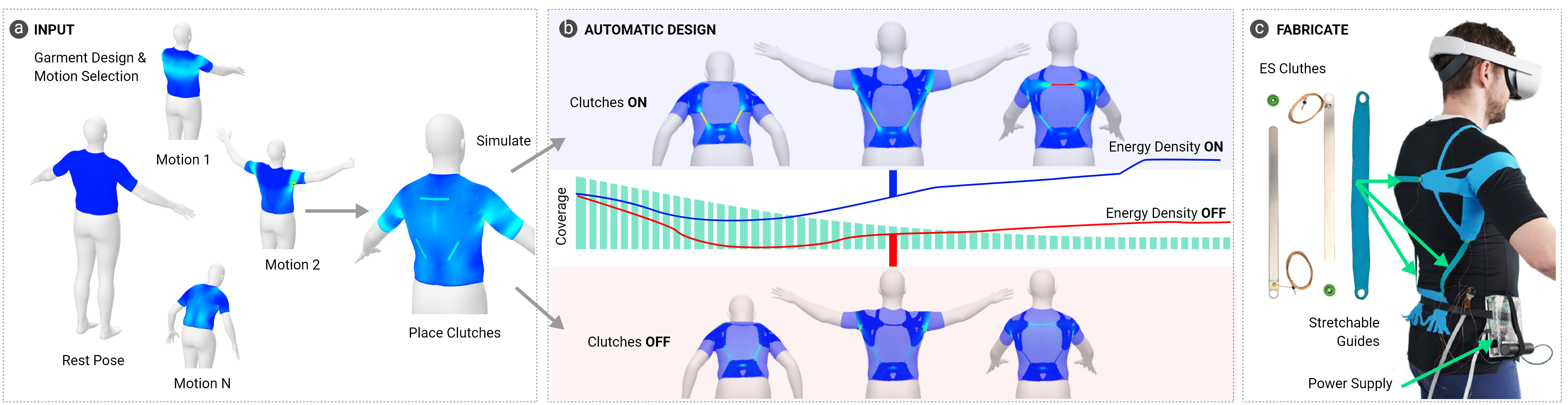}
  \caption{Pipeline overview, from left to right (a) \emph{Input:} Designers specify N motions with a single rest pose, and a garment (shirt). The simulated behaviour of the garment is shown, with light blue indicating high strain. Clutches are then placed on the garment over high strain areas. (b) \emph{Automatic Design:} We simulate the garment under each motion with all clutches ON (top), or OFF (bottom), noting that the energy difference between these states increases  until the target coverage is reached. Designs can be sampled at any point in the evolutionary progress. (c) \emph{Fabricate:} We assemble the ES clutches made of flexible strips sliding in a stretchable textile guide and fabricate the connecting structures attaching everything together to make the final active kinesthetic garment (right).}
  \label{fig-system} \vspace{-.25cm}
\end{figure*}


\section{Computational Design Pipeline}
Our method supports designers in the task of creating \textit{active} kinesthetic garments that can resist any motion from a predefined set of movements. The design goals of our pipeline are to enable kinesthetic garments that maximize the feedback felt by users when ES Clutches are active, while minimizing interference with their motion when inactive. Our pipeline consists of three main phases: 1) input --- where designers specify motions, garment designs, and clutch placements; 2) automatic design --- where our method automatically links clutches with stiff material to satisfy the above design goals; and 3) a fabrication phase. The full computational pipeline is illustrated in Fig \ref{fig-system}. 




\subsection{Input}
Our pipeline requires three components as input: a set of body motion, a base garment, and a predefined number of ES clutches. 

\paragraph*{Motions} are specified using the STAR/SMPL parametric human body model \cite{osman2020star, loper2015smpl} which produces a surface mesh $v$ with N = 6890 vertices $\in\mathbb{R}^{3n}$ based on a $72$ pose  $\mv{\theta}$ parameters. 
To create target poses, we sample from the AMASS dataset \cite{mahmood2019amass} and make individual adjustments to the $\mv{\theta}$ as needed. We define a motion as a single rest pose $\bar{\mathbf{v}}$ and an accompanying deformed pose $\mathbf{v}$. A \emph{set} of motions is defined with $\mathbf{v}=(\mathbf{v}_1,\ldots,\mathbf{v}_i)$ deformed poses, and a common rest pose $\bar{\mathbf{v}}$.

\paragraph*{Garments and Connecting Structures} are modeled as 2D mesh surfaces embedded on the body, initialized with the same rest and deformed nodal positions as the underlying body mesh.  A garment in its rest state is defined through nodal positions  $\bar{\mathbf{x}}=(\bar{\mathbf{x}}_1,\ldots,\bar{\mathbf{x}}_n)\in\mathbb{R}^{3}$ and $\mathbf{x}=(\mathbf{x}_1,\ldots,\mathbf{x}_n)\in\mathbb{R}^{3n}$ when deformed. The connecting structure of the garment is modeled using a bi-material distribution where each triangle element $e$ of the garment mesh is assigned a specific material property. This property is set through the design variable $\materialdensity^e \in [{0, 1}]$ for each element $e$, where $0$ and $1$ correspond to \textit{cloth} and  \textit{reinforced cloth} respectively.

\paragraph*{Active Components}
In our formulation, ES clutches are modeled as rectangular surface meshes that are attached to the garment at a predefined set of vertices. A key requirement for optimal ES clutch operation is that they are initialized in a \emph{taut} state, that is, all slack must be removed from the system before forces are felt at the endpoints. We create a low-dimensional parametrization of ES clutches that is defined by the following variables: a \emph{starting point}, a \emph{surface direction}, and a \emph{length and width}. 
From this, we procedurally generate a spline, and extrude a mesh (see Appendix \ref{ap:procedural_clutch}) with rest vertices $\mathbf{q}=(\mathbf{q}_1,\ldots,\mathbf{q}_m)\in\mathbb{R}^{3n}$ and $\bar{\mathbf{q}}=(\bar{\mathbf{q}}_1,\ldots,\bar{\mathbf{q}}_m)\in\mathbb{R}^{3n}$ when deformed. The endpoints $q^c$ of the clutch mesh (three at each end) are connected to the garment mesh using simple quadratic penalty functions, which allows for firm attachment.

\subsection{Automatic Design}
Finding a passive mechanical structure that optimally connects electrostatic clutches placed by the user is a key challenge in the design of active kinesthetic garments. Recent work by Vechev et al. \cite{vechev2022cdkg}  demonstrated a method for on-body topology optimization using a single compliance-minimization objective (summarized in Appendix \ref{ap:onbody_tpo}). 
Such an objective cannot be applied in our setting, as it has no notion of component states, and the single objective does not sufficiently capture the high-level goal of minimizing motion interference when components are inactive. Therefore, we propose to extend this formulation from passive reinforcements to our setting of active kinesthetic garments by \textit{(1)} distinguishing between active and inactive clutch states by extending the simulation model with stateful components, \textit{(2)} reconciling the different design goals for active and inactive states through a new state-dependant dual-objective, and \textit{(3)} accounting for multiple motions. 



\paragraph*{Active Component Model and Simulation}
ES Clutch stiffness varies according to their state, thus, we model their behaviour using a bi-modal material. We implement this as a neo-Hookean material that resists compression and changes modes depending on the activation vector $\gamma = [\gamma_0, \gamma_1,..., \gamma_n], \quad  \gamma_n \in [{0, 1}]$. Each $\gamma_i$ determines the state of clutch $i$, with $0$ and $1$ corresponding to inactive and active states, respectively. The Young's modulus of the clutch material is then set to $Y_\mathrm{clutch}^i=\gamma_iY_\mathrm{stiff}+(1-\gamma_i)Y_\mathrm{cloth}$. The elastic energy stored in the clutches during deformation is defined as $E_\mathrm{clutches}(q, \gamma)$. We define penalty terms $E_\mathrm{body}(v, \bq)$ preventing clutches from entering the body, and an additional term $E_\mathrm{attach} = \frac{1}{2}k(q^c-x^x_c)^T(q^c-x^x_c)$ that attaches the six endpoint vertices to their respective locations $x_c$ on the garment. 
Throughout all examples, we set constant values for Young's Modulus to $Y_\mathrm{cloth} = 0.5\textrm{MPa}$, $Y_\mathrm{reinforced\_cloth} = 0.5\textrm{GPa}$, and $Y_\mathrm{stiff} = 3.0\textrm{GPa}$. We use a Poisson’s ratio of 0.33 for all materials. 

We combine our active component model, with the garment-on-body model described in \cite{vechev2022cdkg}. The terms $E_\mathrm{garment}(\bx, \materialdensity)$, $E_\mathrm{body}(v, \bx)$, and $E_\mathrm{attach}$ are summarized in Appendix \ref{ap:sim_model}. With the model and energies defined above, we perform a quasi-static simulation by solving an unconstrained optimization problem,

\begin{equation}
\label{eq:CoupledSystemEnergy}
\begin{split}
\bx^*, \bq^* = \arg\min_{\bx, \bq} \quad E_\mathrm{garment}(\bx, \materialdensity) + E_\mathrm{body}(v, \bx) +  E_\mathrm{attach}(\bx) \ + \\  
E_\mathrm{clutches}(\bq, \gamma) + E_\mathrm{body}(v, \bq) +  E_\mathrm{attach}(\bq) \ ,
\end{split}
\end{equation}
using the GPU-based L-BFGS \cite{liu1989limited} optimizer provided by PyTorch \cite{pytorch}. We take advantage of GPU parallelism by simulating all states (poses) simultaneously. We consider simulations converged once the gradient norm of (\ref{eq:CoupledSystemEnergy}) reaches 1e-7.

\paragraph*{State-Dependant Dual-Objective}
A central goal for the structural optimization step is to find a material layout such that the garment resists the specified motions as strongly as possible when clutches are active, while showing minimal resistance otherwise. Assuming all-elastic materials, we translate this goal into the requirement that the stored energy of the garment should be maximized when clutches are active, and minimized when they are inactive. 
Our key insight is to introduce an \emph{energy differential objective} that combines these opposing goals as
\begin{equation}
\label{eq:SingleMotionObjective}
\begin{split}\mathbf{d}^* = \arg\max_{\mathbf{d}}  \quad E_\mathrm{garment}(x_{ON}^*(\mathbf{d, q, \gamma}), \mathbf{d}) \\
- E_\mathrm{garment}(x_{OFF}^*(\mathbf{d, q, \gamma}), \mathbf{d}) \\ \textrm{s.t.} \quad \sum_e{A_e} d_e=A^*\ ,\quad \mathbf{f}(x_{ON}^*)= \mathbf{0},  \quad \mathbf{f}(x_{OFF}^*)= \mathbf{0} \end{split} \ ,
\end{equation}
where $\bq$ holds the variables of all clutches, and $x_{ON}^*$, $x_{OFF}^*$  are distinct equilibrium states  corresponding to all clutches being active ($\gamma_i=1 \forall i$) and inactive ($\gamma_i=0 \forall i$), respectively. 

To solve this optimization problem with the BESO algorithm, we must compute the per-element sensitivities, i.e., the partial derivatives of the objective function with respect to per-element material assignment variables $d^e$. Following (\ref{eq:SingleMotionObjective}), we simply have to sum the sensitivity values for the active and inactive states to obtain a single value that is used in the BESO ranking procedure. Everything else follows the procedure described in \cite{vechev2022cdkg} and is summarized in Appendix \ref{ap:onbody_tpo}.

\paragraph*{Multiple Motions}
Whereas the method described in \cite{vechev2022cdkg} computes static reinforcements for a single target motion, we ultimately want to move towards \emph{programmable garments} that can resist many motions by use of their active components. 
To this end, we extend (\ref{eq:SingleMotionObjective}) to the multi-motion setting by summing contributions for all poses as
\begin{equation}
\label{eq:MultiMotionObjective}
\begin{split}\mathbf{d}^* = \arg\max_{\mathbf{d}}  \quad 
\sum_k \hat{E}_\mathrm{garment}^k(x_{k, ON}^*(\mathbf{d, q, \gamma}), \mathbf{d}) \\ 
- \sum_k \hat{E}_\mathrm{garment}^k(x_{k, OFF}^*(\mathbf{d, q, \gamma}), \mathbf{d})   \\ 
\textrm{s.t.} \quad \sum_e{A^e} d^e=A^*\ ,\quad \mathbf{f}(x_{k, ON}^*)= \mathbf{0},  \quad \mathbf{f}(x_{k, OFF}^*)= \mathbf{0} \ \forall k \ ,
\end{split}
\end{equation}
where $k$ runs over all input poses.
A problem with this simple approach is that the optimization may receive larger rewards for increasing an already good performance for a given pose instead of improving the worst-performing case.  
We address this problem by normalizing the strain energy density for each pose in a pre-processing step
\begin{equation}
\label{eq:NormalizedGarmentEnergy}
\hat{E}_{\mathrm{garment}}^k = \sum_e t^e  A^e\hat{\energydensity}_\mathrm{garment}^{k,e}(x^*, \materialdensity^e), \ \hat{\energydensity}^{k,e} = \frac{\energydensity^{k,e}}{\max_e(\energydensity^{k,e})} \ .
\end{equation}
In this way, each pose is given the same importance, irrespective of its initial strain energy, thus encouraging material layouts that more evenly distribute the garment's performance across all input motions.




\subsection{Hardware Details and Fabrication}
In the last step of the pipeline, designs are fabricated. 

\paragraph*{ES Clutches} provide resistance to elongation when active \cite{hinchet2019highforce}, while remaining stretchable with low resistance when inactive. They are thin, light and flexible which make them highly compliant and consume very low power when engaged (e.g. one 14cm by 1cm clutch consumes 12 mW at 350V). The ES clutches from \cite{hinchet2019highforce} were modified for better integration by making them stiffer to reduce bending, packaging them in elastic guides to keep them fully self-retractable and safer for on-body use. Each ES clutch is composed of 3 parts: an electrode strip, an insulating strip, and a stretchable textile guide. Strips are made of 125 $\mu$m metalized polyester films from McMaster-Carr. Films are laser cut into long 1cm wide strips of various lengths. Additionally, insulating strips are covered with a 25$\mu$m thick layer of poly (vinylidene fluoride-trifluoroethylene-chlorotrifluoroethylene) from Piezotech-Arkema \cite{hinchet2019highforce}. 

\paragraph*{Garments and Attachments}
All designs are exported as meshes and manually processed in Blender. We simplify geometry, and unroll the designs onto flat surfaces using the Paper model plugin (without changing area). As our connecting material, we attached a layer of polyurethane (Siser EasyWeed Stretch) onto 100\% cotton fabric. This material combination enables much higher forces than in \cite{vechev2022cdkg}. As base garments we used stretchable GripGrab UV sleeves and Nike Dri-Fit Pro Compression shirts. The different parts of the connecting structure were cut with a Trotec 300 laser cutter and glued onto base garments following marks taken on an experimenter wearing the garment. Next, pressure buttons are riveted at locations where ES clutches connect. Finally, ES clutches are fixed onto the garments using pressure buttons and connected with thin wires to a custom voltage power supply powered by a USB power bank and controlled by Bluetooth (see Fig. \ref{fig-system}b). The overall modular system can accommodate different sizes of clutches and slight variations in body sizes.



\section{Evaluation}
We conduct a multi-faceted evaluation of our method showing results for different types of motions and garments in simulation, a mechanical force study, and a VR pointing task. 



\subsection{Automatic Designs}
\begin{figure}[h]
 \center
  \includegraphics[width=1.0\columnwidth]{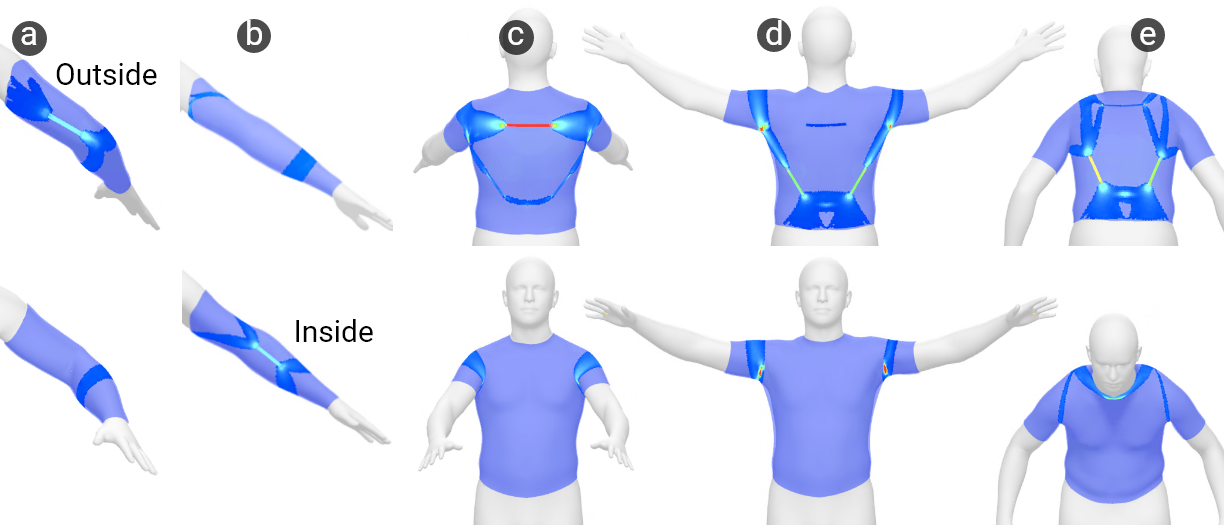}
  \caption{Single-Motion designs for (a) Arm Flexion, (b) Arm Extension, (c) Arms Forwards, (d) Arms Raise, and (e) Bend Forwards. Color coding indicates energy density.  }
  \label{fig-single-motions} \vspace{-.25cm}
\end{figure}

\begin{figure}[h]
 \center
  \includegraphics[width=1.0\columnwidth]{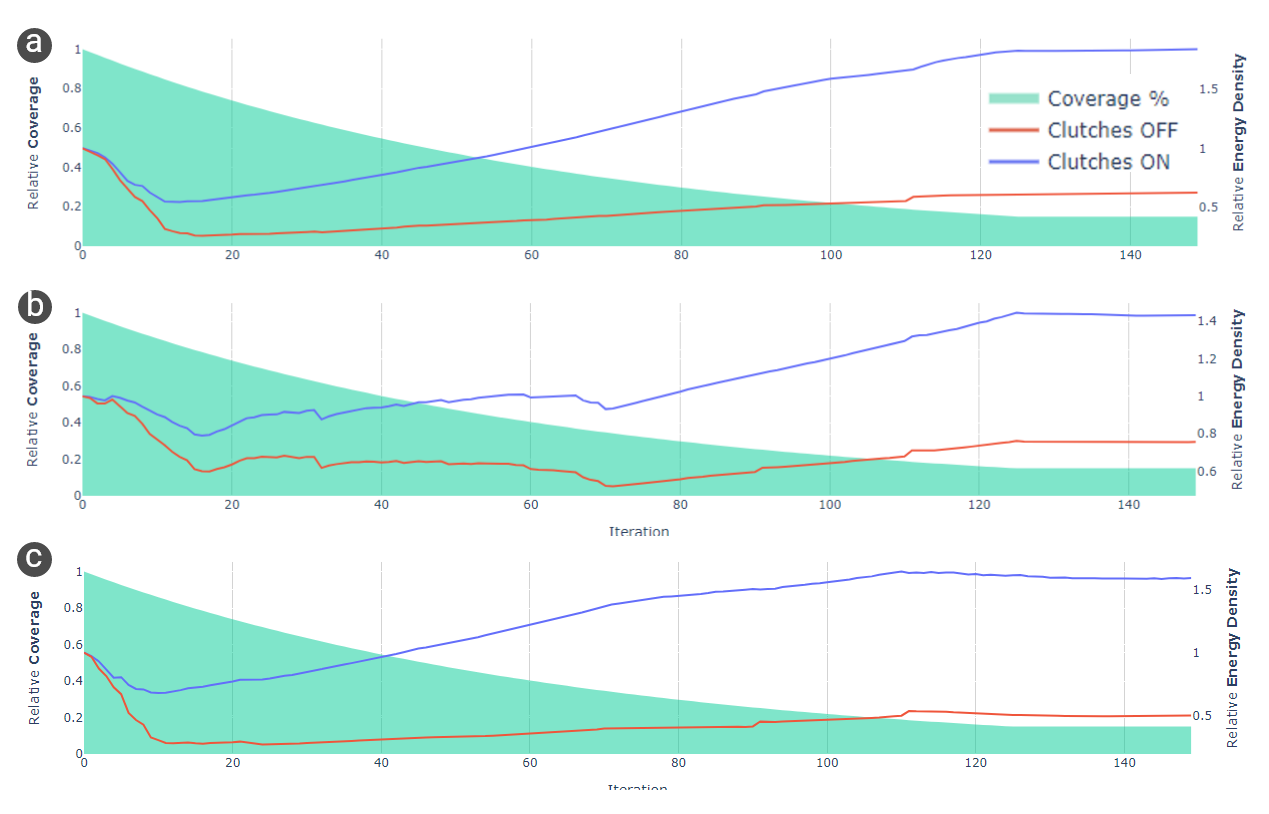}
  \caption{Progress of the evolutionary optimization algorithm for the Shirt Design for (a) Arms Forwards, (b) Arms Raise, and (c) Bend Forwards. }
  \label{fig-evolutionary-progress} \vspace{-.25cm}
\end{figure}

We show a range of designs produced by our method for a variety of motions and garment types. For all experiments, we use a common rest pose with the body in an A-pose, and sample from a set of motions that include \emph{Arms Forward, Arms Raise, Arm Flexion, Arm Extension, Bend Forwards} (see Fig.\ref{fig-single-motions}). Three garments are designed using our tool to cover a variety of body sites: a short-sleeve shirt, an arm-sleeve, and a long-sleeve shirt. All clutches are placed manually on the garments, typically over high strain energy areas of the garments (see Fig. \ref{fig-system}a).
We set the following standard BESO parameters for all experiments: evolutionary rate $ER = 1.5\%$, maximum material added per iteration $AR = 1.5\%$, material interpolation $p=1.6$. Similarly, we set the material budget to $A^*= 15\%$ for all examples except for arm flexion and extension, where we use $A^*$ = 20\%. As our primary metric, we use the \emph{relative energy density}  
\begin{equation}
\rho(\gamma,\mathbf{d}) =  \frac{E_\mathrm{garment}(x^*(\gamma,\mathbf{d}),\mathbf{d})\cdot A_\mathrm{dense}}
{E_\mathrm{garment}(x^*(\gamma,\mathbf{1}),\mathbf{1})\cdot A_\mathrm{opt}} \ ,
\end{equation}
i.e., the ratio between energy density for the optimized and fully dense designs.
As the optimization progresses, we expect to see a widening gap in this metric between active and inactive clutch states (see Fig. \ref{fig-evolutionary-progress} for a visualization).

\paragraph*{Single-Motion Designs}
We begin by showing results for the single motion cases of our method. We target Arm Flexion with a single (8cm) clutch on the elbow, and Arm Extension also with a single (8cm) clutch on the inside of the forearm. We show two separate results in Fig. \ref{fig-single-motions} a, and b. Relative to the fully dense design, we see that energy density \emph{increases} to 1.14 for flexion, and 1.74 for extension when clutches are active. When clutches are inactive, relative energy density \emph{decreases} to 0.34 and 0.54, respectively.

Next, we target single motions on the upper body using three clutches of 15cm length.  Fig. \ref{fig-single-motions} shows results for Arms Forward, Arms Raise, and Bend Forwards, with increases in relative energy density of 2.13, 1.51, and 2.47 respectively. For deactivated clutches, we observe that relative energy density decreases to 0.48, 0.66, and 0.73 for each design.

\begin{figure}[h]
 \center
  \includegraphics[width=1.0\columnwidth]{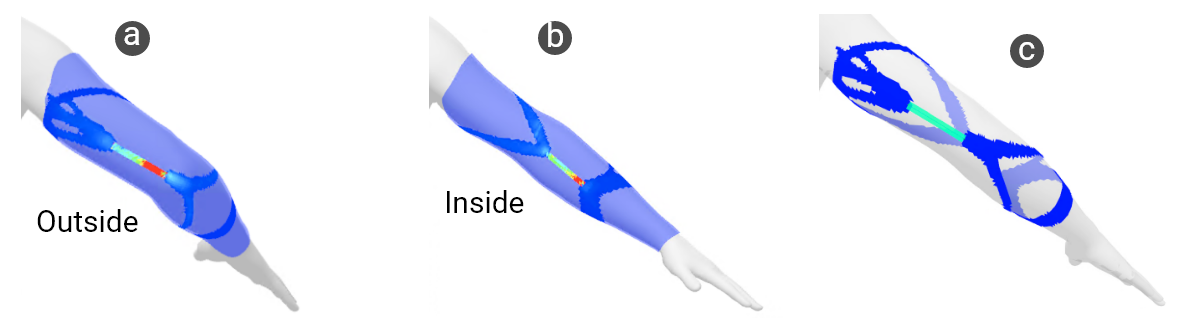}
  \caption{Multi-motion design for simultaneously optimized for (a) Arm Flexion and (b) Arm Extension. This design effectively integrates the single motion designs of Fig. \ref{fig-single-motions} into an intertwined structure (c). }
  \label{fig-multi-motion-sleeve} \vspace{-.25cm}
\end{figure}

\paragraph*{Multi-Motion Designs}
The ability to resist multiple motions with a single design is an important step towards programmable active kinesthetic garments. We used our method to create three such designs, starting with an arm sleeve design (Fig. \ref{fig-multi-motion-sleeve}) that combines Flexion and Extension. It uses the same 20\% material budget as in the single motion designs, but now this material must be distributed to balance performance for two distinct motions. The optimized design achieves relative energy densities of 0.88 and 1.27 for Flexion and Extension, respectively, which is 77\% and 73\% of the corresponding single-motion designs. For perspective, when evaluating the single-motion designs for Flexion/Extension on the Extension/Flexion motion, the relative efficiency is only 2\%/5\%. These results are not unexpected as Flexion and Extension are orthogonal motions  such that designs optimized for only one of them are ineffective for the other one.

\begin{figure}[h]
 \center
  \includegraphics[width=1.0\columnwidth]{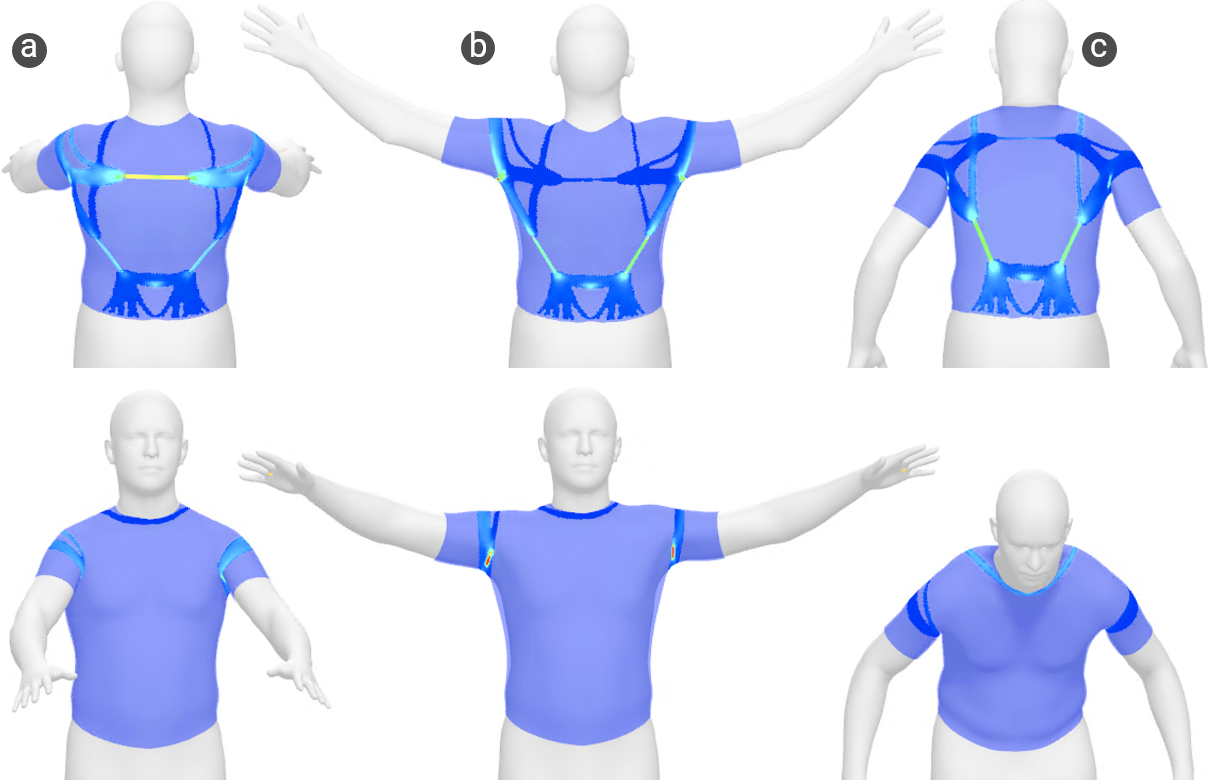}
  \caption{Active kinesthetic shirt designed for the three motions: (a) Arms Forwards, (b) Arms Raise, and (c) Bend Forwards. Strain energy density is shown in color-coding with increasing intensity from \textit{dark blue} to \textit{red}.}
  \label{fig-multi-motion-shirt} \vspace{-.25cm}
\end{figure}

Our second design is a shirt that combines three upper body motions as shown in Fig. \ref{fig-multi-motion-shirt}. 
Many of the features observed in the single-motion versions can be seen here, with clutches linking disconnected reinforcements. It is worth noting that each of these motions leads to a distinct load path (light green/yellow) running through at least one of the clutches. We also compare the performance of the multi-motion design to the single-motion versions. As can be seen in  Table. \ref{tbl:shirt-designs-comparison}, the multi-motion design is within 83\%, 72\%, and 65\% as efficient as the single-motion designs, and yet using the same material budget. The performance of the single-motion designs on  motions for which they were not optimized is, again, significantly lower.

Additionally, for each motion we show the progress plots of the evolutionary optimization in Fig. \ref{fig-evolutionary-progress}. As our automatic design method removes material, we see a clear separation in relative energy density for active and inactive states for all three motions. In the \emph{inactive} mode, the relative energy densities of the garment for each motion are \emph{decreased} by 0.62, 0.72, and 0.5, showing that our method is able to consistently achieve its minimization objective.


\begin{table}[h]
\begin{tabular}{ p{2.75cm} p{1.0cm} p{1.0cm} p{1.0cm} p{1.0cm} }
\hline
\multirow{2}{*}{ Evaluated On}  & \multicolumn{4}{c}{Optimized For } \\
& Forwards & Raise & Bend & All   \\ \hline
\rule{0pt}{3ex}Arms Forwards &  2.13 & 0.46 & 0.34 & 1.77 \\ 
Arms Raise    &  0.62  & 1.51 & 0.14 & 1.08\\ 
Bend Forwards    &  0.35  & 0.81 & 2.47 & 1.60\\ \hline

\end{tabular}
\caption{Comparison of garments optimized for a single motion against a garment optimized for all three motions. A higher number corresponds to an increase in relative energy density when clutches are active.}
\label{tbl:shirt-designs-comparison}
\end{table}

Our final example investigates the scalability of our method to more complex scenarios involving five clutches and five motions. The performance of this design exhibits relative energy density increases of 1.41, 0.85, 1.62, 0.64, and 1.16 for the motions Arms Forwards, Arms Raise, Bend Forwards, Arm Flexion, and Arm Extension, respectively. These numbers are comparable to the results obtained for our other multi-motion garments, especially as the allotted per-motion coverage has decreased overall. In general, the more motions a given design supports with the \emph{same} material coverage target (i.e. 15\%), the material available per motion will decrease and thus be less energy-dense in the ON state. In this case, the material coverage target can be increased, or the designer can sample from an earlier progression step with higher coverage. 



\begin{figure}[h]
 \center
  \includegraphics[width=1.0\columnwidth]{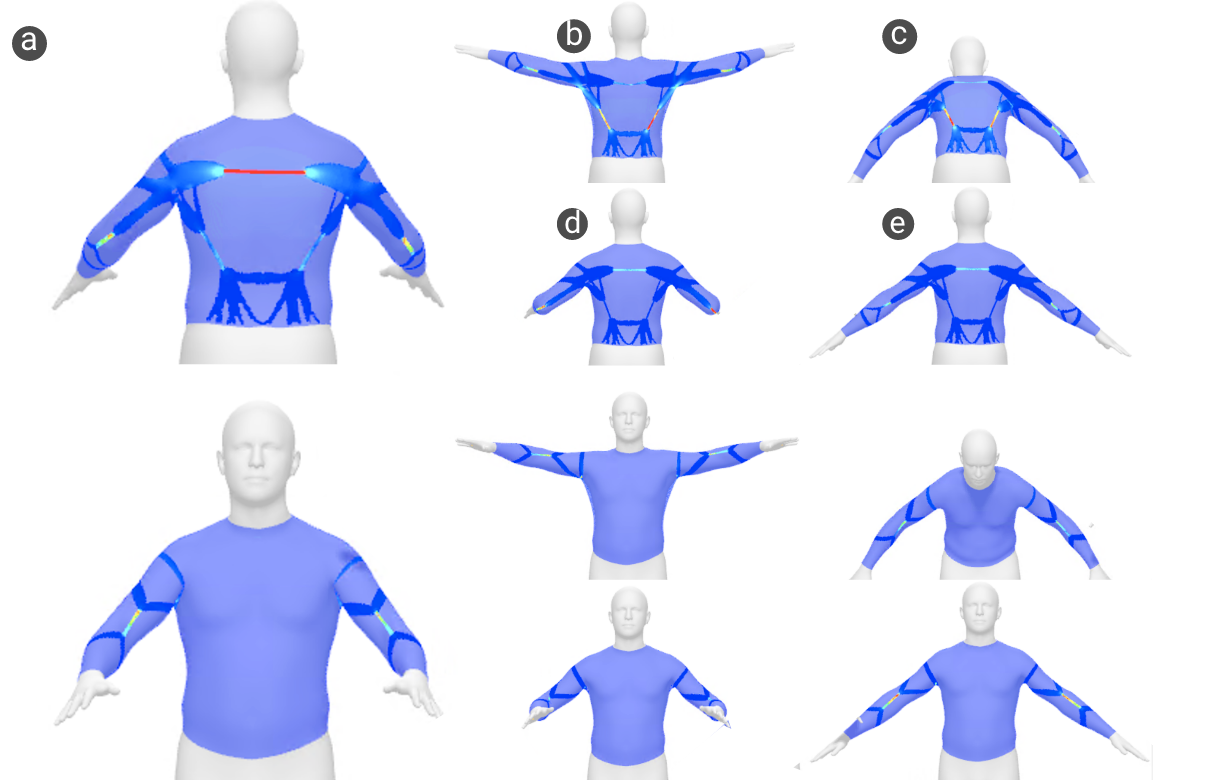}
  \caption{Active kinesthetic long-sleeve shirt with five clutches designed for five motions as indicated.}
  \label{fig-multi-motion-long-sleeves} \vspace{-.25cm}
\end{figure}

\subsection{Comparison to Manual Designs}
We conducted a pilot study to provide a manual baseline for our automatically generated designs. A central question in this context is whether users converge towards particular designs and if those designs exhibit features found in automatically generated ones. We recruited six participants (5M, 1F), two of whom were experts in structural optimization techniques (P2, P3). Using our interactive tool, we asked users to 'draw' stiff material on garment meshes, connecting a set of already placed clutches. Participants were asked to distribute material in such a way as to maximally resist the set of specified motions when clutches are activated. Each participant created two designs, a 2-clutch, 2-motion arm sleeve, covering no more than 20\% of the available area, and a 3-clutch 3-motion shirt with a coverage budget of 15\%. Each of these designs corresponds to an automatically generated designs shown in the previous section. The secondary goal of minimizing energy when the clutch is inactive was not assigned. 

\begin{figure}[h]
 \center
  \includegraphics[width=1.0\columnwidth]{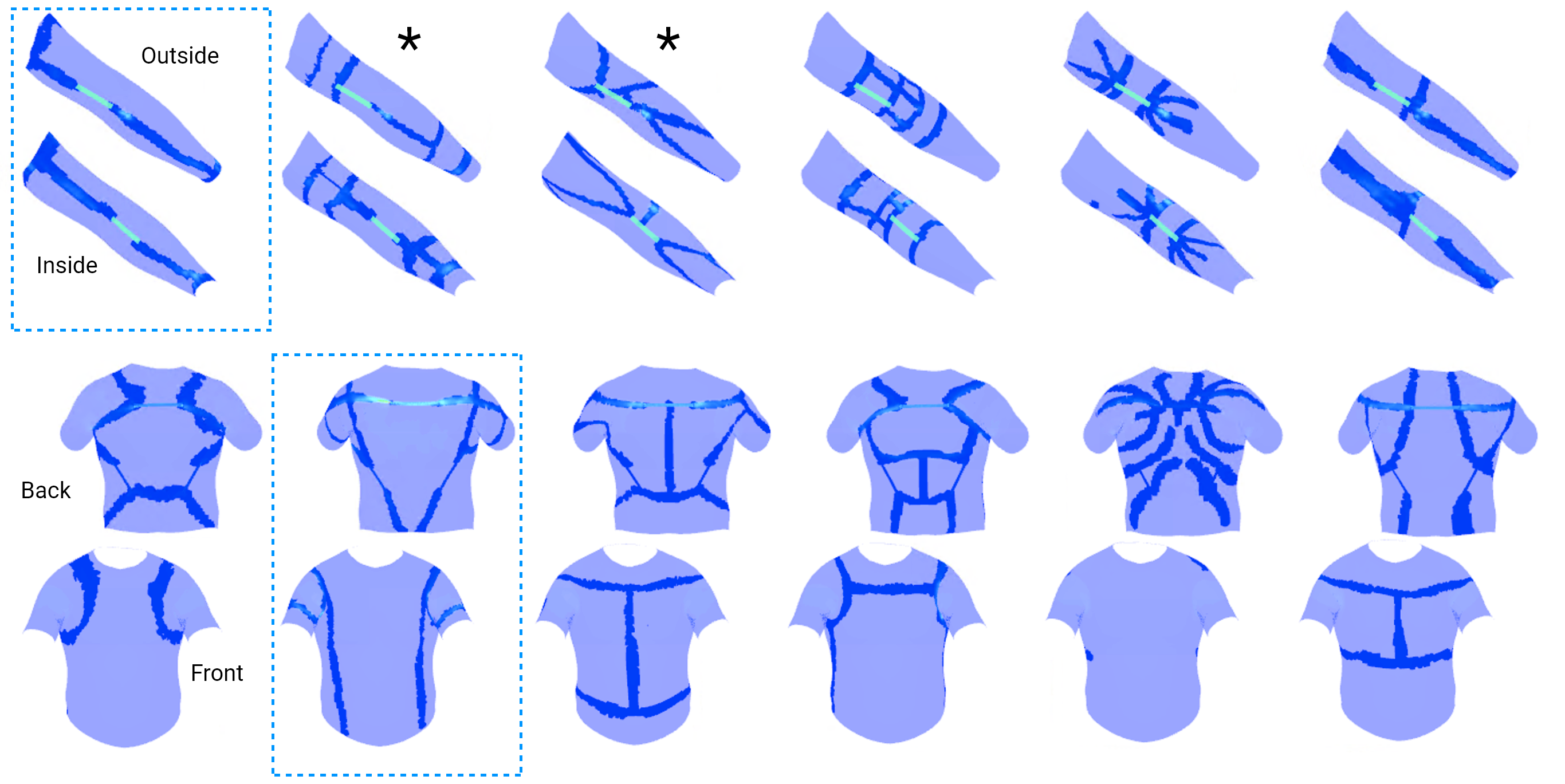}
  \caption{Manually-designed garments for 2-motion arm sleeve (top), and 3-motion shirt (bottom) for participants P1 (left) to P6 (right) with clutches shown in green. Note the large variance among the designs, particularly in the shirt case. The sleeve from P1 and the shirt from P2 were chosen for fabrication. * Denotes expert designed garments.}
  \label{fig-manual-designed-garments} \vspace{-.25cm}
\end{figure}

The manually-created results shown in Fig. \ref{fig-manual-designed-garments} exhibit large variety in their designs.
While most examples can be expected to perform reasonably, none of them resembled the automatically generated designs. Compared to the fully dense version, manually-created designs were only 0.48x and 0.27x as energy-dense for the arm sleeve and shirt, respectively. Automatic designs, on the other hand, showed a 1.1x and 1.48x higher energy density. We can see that in the case of designing for a larger number of motions, the effectiveness of user designs drops drastically, while automatically generated designs can maintain a relatively high energy density. Looking at only designs from expert users, we see relative average energy densities of 0.54 for the sleeve, and 0.34 for the shirt, still much lower than our automatic designs. Non-expert designs on the other hand had average relative energy densities of 0.44 for the sleeve and 0.23 for the shirt, showing a much larger drop in performance for the more complex shirt design. Thus, automatic design methods can be especially useful for such users. Table \ref{tbl:manual-designs} summarizes these findings. 

\begin{table}[h]
\begin{tabular}{ p{2.0cm} p{0.5cm} p{0.5cm} p{0.5cm} p{0.5cm} p{0.5cm} p{0.5cm} p{0.5cm} }
\hline
Garment / \\Motion   & Auto & P1 & P2$^*$  & P3$^*$  & P4  & P5 & P6 \\ \hline
\rule{0pt}{3ex}Sleeve / Flex & \textbf{0.88} & 0.50 & 0.53 & 0.54 & 0.67 & 0.26 & 0.58 \\
Sleeve / Ext & \textbf{1.28} & 0.37 & 0.66  & 0.44 & .55 & 0.21 & 0.45 \\
Shirt / Forward & \textbf{1.77} & 0.11 & 0.45 & 0.14 & .19 & 0.08 & 0.14 \\
Shirt / Raise & \textbf{1.08} & 0.10 & 0.37  & 0.07 & .14 & 0.02 & 0.19 \\
Shirt / Bend & \textbf{1.60} & 0.38 & 0.75  & 0.27 & .64 & 0.32 & 0.47 \\

\hline
\end{tabular}
\caption{Performance summary of manually-created designs. We report the energy density of the garment relative to the fully dense design. Note that the automatic design has the highest energy densities across all motions. * Denotes designs by expert users.}
\label{tbl:manual-designs}
\end{table}



\subsection{Physical Validation}
We seek to quantify the resistive force of our automatically designed garments under the motions for which they were optimized, and compare them against Manual-Design counterparts. We selected two designs for fabrication - the multi-motion arm sleeve and multi-motion short-sleeve shirt. We fabricated both, the designs produced by our automatic method and the corresponding manually-designed garments. For the shirt, we selected the clearly highest performing garment, which was from P2, while for the sleeve, we selected the design from P1. This sleeve design represents a common (line) design seen in literature \cite{Ramachandran2021, ramachandran2021arm, Diller2016}, while having similar performance as other designs.



\begin{figure}[h]
 \center
  \includegraphics[width=1.0\columnwidth]{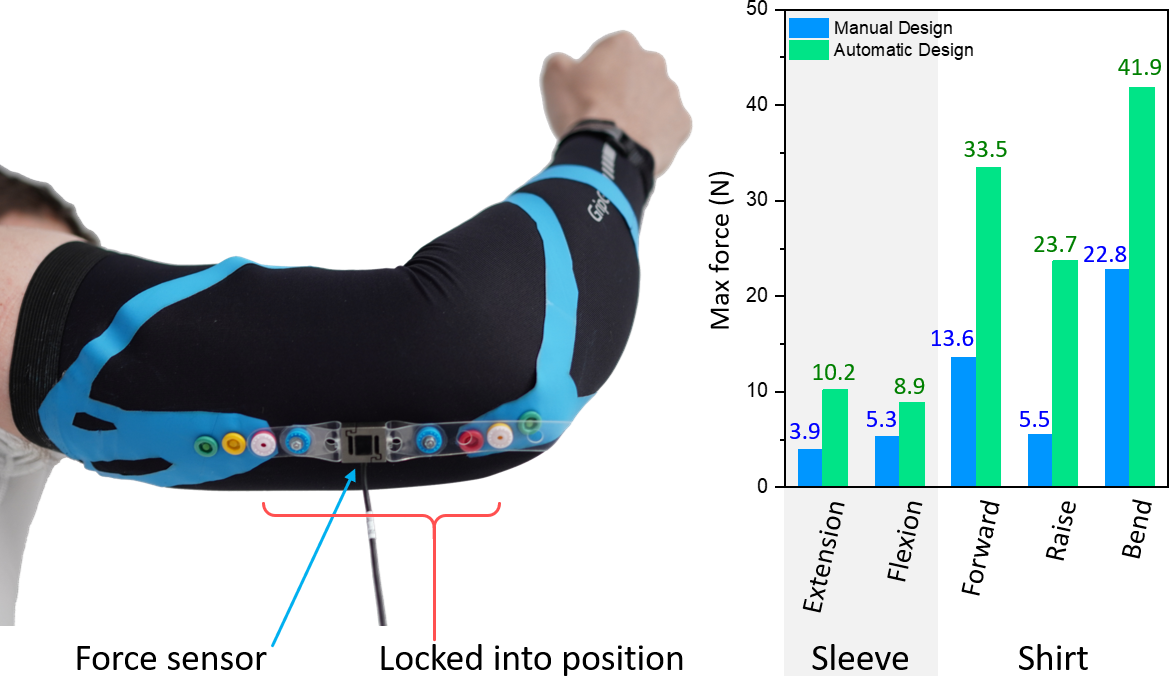}
  \caption{Physical force measurements. Left: experimental setup with force with clutch replaced by a stand-in equipped with a force sensor. Right: maximum force (N) readings when blocking different movements as labeled for the manually-designed (blue) and automatically generated (green) shirt and sleeve garments.}
  \label{fig:force_graph} \vspace{-.25cm}
\end{figure}

In order to best isolate the impact of the connecting structure, we replace clutches with flexible plastic strips that connect to a force sensor as shown in Fig. \ref{fig:force_graph}. For the Arms Forward motion, we mount the force sensor in the upper back, while for the Arms Raise and Bend Forward motions, we mount it on the bottom left clutch location. The target motion is then slowly performed by the experimenter wearing the garment (three  trials per motion), while the force is measured using a 10kg DYLY-108 force sensor with an HX711 load cell amplifier (see the Video Figure for visual demonstration).
The results shown in Fig. \ref{fig:force_graph} indicate that, relative to the manual design, the designs generated by our method were on average two times and up to four times more efficient in terms of force output. These measurements confirm our observations made on simulation results in which, as for the experimental case, the largest difference in relative energy density was observed for the Arms Raise motion.

\subsection{User Evaluation}

\label{sec:eval-user}
To quantitatively evaluate the ability of our active kinesthetic garments to efficiently block motion, we conduct a user study based on a VR pointing task in which participants were asked to reach targets from a predefined set of locations within their reach.
The hypothesis that we seek to test is that, when wearing our optimized designs, users generally need more time to reach targets when clutches are active compared to when they are inactive. A secondary hypothesis is that our automatically generated designs lead to higher blocking efficiency than a user-generated baseline.


\paragraph*{Procedure and Setup}
Six healthy adult subjects ($M$=28.1; $SD$=4.14;) were recruited. Since we only fabricated one size of our designs, participants were all male and similar in size to the template STAR mesh. All participants wore noise cancelling headphones. The procedure and tasks were described and an introduction to the garments and the active components was given. After donning the garments (shirt and sleeve), clutches were attached and adjusted according to participant size to achieve sufficient pre-tension. The left hand of participants was rested on a tripod such for stability. 
Participants were then introduced to the VR setting and asked to practice touching the spherical targets with and without clutch activation until they felt comfortable proceeding. The study was implemented in Unity 2021 using a Meta Quest 2 relying on the built-in hand-tracking functionality. 

\begin{figure}[h]
 \center
  \includegraphics[width=1.0\columnwidth]{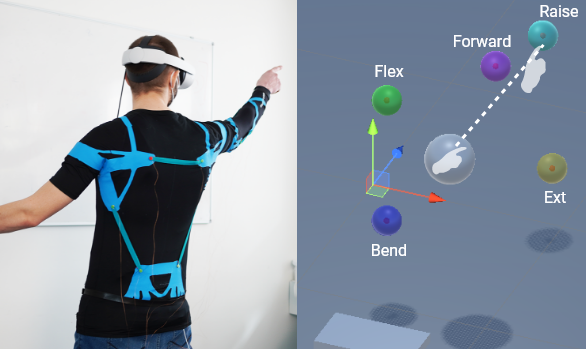}
  \caption{User study setup. Participant wearing Auto-Designed garment and reaching for target (left), and their corresponding motion in the virtual environment (right).}
  \label{fig:study_setup} \vspace{-.25cm}
\end{figure}

\paragraph*{Study Design}
We use a within-subject design with two independent variables: \emph{Feedback Type \{via Auto-Designed, via Manual-Designed, Visual Only\}} and \emph{ Target Placement: \{Forward, Raise, Bend, Flex, Ext\}}. Each target is placed to elicit a specific motion from the user, and is color-coded to 4 to participants which target they should touch (see Fig.\ref{fig:study_setup}). As a dependent variable, we measure \emph{Time}, which starts automatically when the participant's hand leaves the starting position (white sphere), and ends as soon as they touch it again. The main task was to touch a given target in one continuous ballistic back-and-forth motion at a natural speed. For each target placement, three trials were collected for a total of 30 trials, one half with clutches active, the other half with clutches deactivated (Visual). The order of clutch activation  was randomized and participants were not told if the clutch was on or off. The order of the feedback type was also randomized. At the end of the study, participants were free to comment on their experience using each garment design.

\begin{figure}[h]
 \center
  \includegraphics[width=1.0\columnwidth]{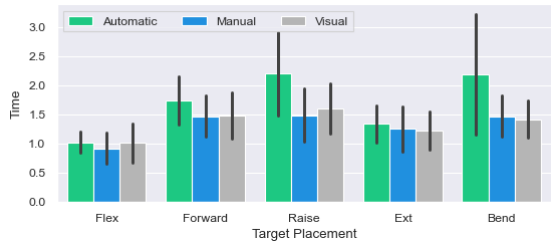}
  \caption{User study results showing the average trial time for each feedback type and target location. }
  \label{fig:study_results} \vspace{-0.25cm}
\end{figure}

\paragraph*{Results}
The mean time to reach targets were 1.68s ($\sigma=0.77)$ for the Auto-Designed condition, 1.32s ($\sigma=0.43)$ for Manual-Designed, and 1.33s ($\sigma=0.43)$ for Visual. A longer reach time indicates more impact on the participant's ability to reach the target. The full results are visualized in Figure \ref{fig:study_results}. A two-way repeated-measures ANOVA resulted in a significant effect on feedback type ($F(2, 5) = 29.82,\  p<.001$), target placement ($F(4, 5) = 34.45, \ p<.001$) and interaction ($F(8, 5) = 5.72, \ p=.004$). We conducted a Holm-corrected post-hoc test and found significant differences for feedback type. Our Automatic Design method significantly impacted participant movement time compared to both Manual Design feedback ($p<.001$) and Visual feedback ($p<.001$). We found no significant difference between Manual Design feedback and Visual feedback.  When looking at times across target placements, our Automatic Design method significantly impacted participant movement time for the Bend and Raise motions when comparing to both Visual and Manual Design baselines (both $p<.001$).



From these results, we see a trend that the automatically designed garments performed better in terms of limiting user motion, particularly when the motions involved larger movements in the upper body. 

We observe that our Auto-Designed garments performed substantially better in larger motions than the Manual-Design counterparts, results which are in-line with both simulated and force-characterization data.

The exception is the Forward motion, where we observed a less substantial impact, possibly due to the fact that participants could twist their body to reach that target. The low performance of the Flex and Ext methods could be due to the fact that we use smaller ES clutches for these motions, and the force was too small compared to the force produced by larger motions (see Fig. \ref{fig:force_graph}), and thus, below a critical threshold that would have an impact on user motion. Thus, our first hypothesis was confirmed for two of the three larger upper body motions.

What is surprising is that the performance of the Manual-Design baseline was nearly indistinguishable from the Visual baseline, even for larger motions. In relation to this, two participants commented that they had trouble perceiving any resisting effects of the Manual-Design.This shows that, even with the same active components, our optimization-based approach for designing connecting structures can indeed make the difference between a system having clear or negligible impact on user motion.

\section{Example Applications}
We show four applications enabled by the ability of active kinesthetic garments to selectively and dynamically engage clutches with a single design.



\begin{figure}[h]
 \center
  \includegraphics[width=1.0\columnwidth]{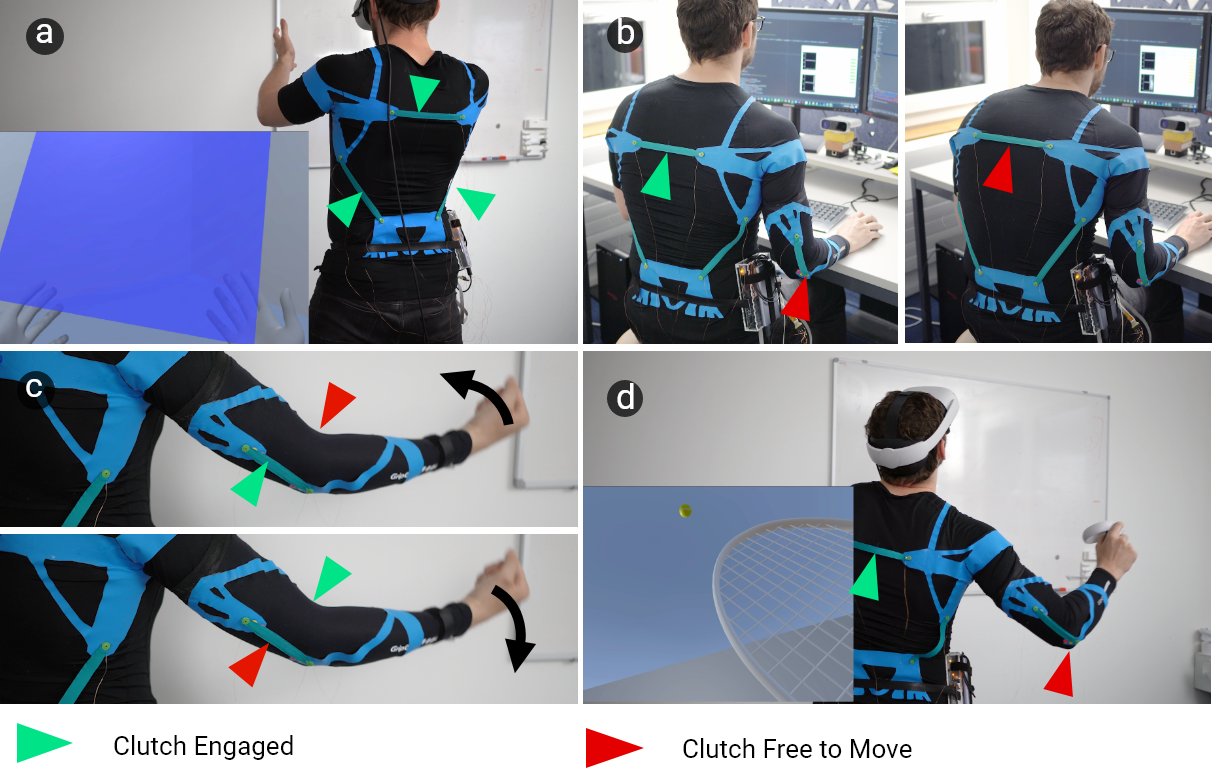}
  \caption{Applications in (a) Workplace Training, (b) Posture Correction, (c) Resistance Training, and (d) VR Gaming.}
  \label{fig:applications} \vspace{-.25cm}
\end{figure}

\paragraph*{Workplace Training}
When picking up a virtual box, we activate all ES clutches on the shirt to provide stability to the upper-body, preventing the arms from going through the box, and preventing the user from overly bending their back. More complex motion control could also provide further training and guidance in combination with a complex control loop (i.e. using body pose as an input).

\paragraph*{Posture Correction}
Bad posture is a very common problem when sitting at a desk, and many posture correcting shirts already exist to help this issue. However, only active kinesthetic garments can periodically allow the user to go into a slouching posture on demand, in addition to keeping other limbs completely unrestricted (i.e. elbow).

\paragraph*{Resistance Training}
Multiple clutches can be selectively activated to resist a target motion, the upper clutch in the case of arm extension, and the lower clutch in the case of arm flexion. The opposing clutch is meanwhile disabled, to prevent full arm-locking. This shows how a single garment can be re-configured at run-time for resisting multiple motions, potentially encompassing a user's entire workout.

\paragraph*{VR Gaming}
VR immersion can be increased significantly by providing physical forces when users make contact with the world. In this game, a user practices hitting tennis balls out of the air, and only the upper back clutch is activated on contact, noting that the elbow clutch remains off and does not prevent natural elbow bending during such sports movements.


\section{Discussion and Future Work}

Our user study results indicate that automatically-designed active kinesthetic garments were able to have a significant impact on user motion, whereas the manually designed counterparts could not meet this threshold, indicating the need for automated methods to assist designers in such tasks. 





\begin{figure}[h]
 \center
  \includegraphics[width=1.0\columnwidth]{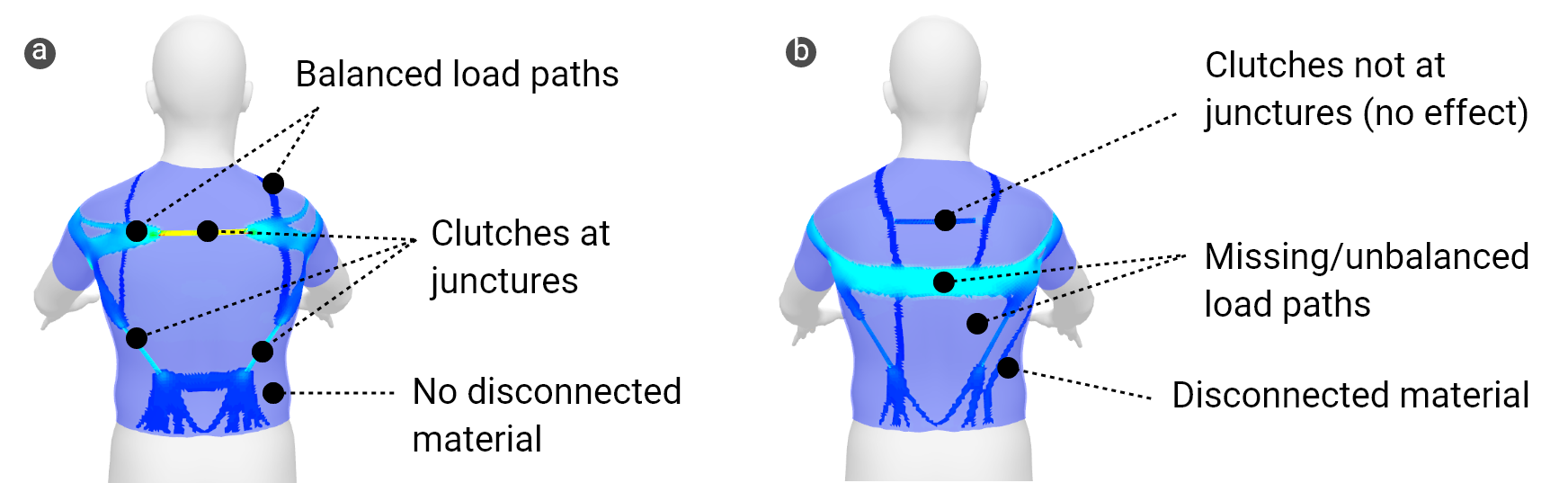}
  \caption{Emergent structural properties for the multi-motion shirt designed with dual-objective minimization (a) vs single compliance minimization from  \cite{vechev2022cdkg} (b).}
  \label{fig:structural_properties} \vspace{-.25cm}
\end{figure}

\paragraph{Emergent Structural Properties}
We found in our evaluation three emergent structural properties: 1) no connecting material is isolated from the main structure (no disparate island) 2) all active components are at junctures of connecting material, and 3) overlapping, yet distinct load paths are created for each specific motion. When comparing designs using our dual-objective directly to the single compliance minimization objective in \cite{vechev2022cdkg} (b), we find that these same properties do not emerge (See Fig. \ref{fig:structural_properties}). Each property plays an important role --- for example, if clutches are not at junctures, then their activation will have no effect on the user. Similarly, unbalanced load paths and islands of disconnected material may degrade performance for particular motions and comfort respectively. Users on the other hand performed well in terms of connecting clutches, but struggled to balance load paths, leading to very poor performance in particular motions.

\paragraph{Limitations}
The main work of the designer in our tool is in the manual placement of ES clutches. As the number of active components and motions grows, the requirement for manual clutch placement may become more and more challenging. Our method can be extended to optimize for ES clutch placement, thereby freeing designers from this task, potentially increasing the relative efficiency of the design. Our method is also limited to a simple mode of activation, where clutches are either all active or inactive. However, clutch activations can be controlled individually and continuously through voltage input that affect the degree of resistance. Accounting for these degrees of freedom during design could further improve efficiency and allow for more targeted resistance to selected motions.

While our active kinesthetic garments are fully wearable and mobile, they do not have any sensing capabilities. Integrating sensing could be done via capacitive sensors, which could be optimized based on the same strain-maximization principle used for connecting clutches. 

Our study was limited to only six male participants and the type of feedback collected was mainly quantitative. Our method could be used to create garments for female users and even \emph{personalized} garments by simply changing the $\beta$ and gender parameters in the SMPL/STAR models. Richer VR interaction opportunities can be explored in the future by moving beyond simple button presses and object intersections, for example, by integrating body-pose sensing into the control loop.



\section{Conclusion}
We presented a computational approach for automatic design of active kinesthetic garments that block user-specified body motion on demand. As our core technical contribution, we cast the design of reinforcing structures that connect and anchor individual clutches as an on-body topology optimization problem and introduced a novel objective term that encourages maximum resistance of the garment when clutches are active while minimizing interference with body motion when they are inactive. Our experiments indicate that our designs are highly effective and consistently and significantly outperform user-created designs.

The structure optimization techniques developed here have the potential to be useful in the routing and placement of other types of active components such as actuators and sensors. By laying out a theoretical and algorithmic basis for this central problem, we hope that our work will serve as a step toward computational design of highly integrated multi-modal wearable interfaces in the future.

\begin{acks}
This work was supported in part
by grants from the Hasler Foundation (Switzerland) and funding from the European Research Council (ERC) under the European Union’s Horizon 2020 research and innovation programme grant agreement No 717054.
\end{acks}

\bibliographystyle{ACM-Reference-Format}
\bibliography{references}

\section*{Appendix}
\appendix

\section{Procedural Clutch Generation}
\label{ap:procedural_clutch}


ES clutches are defined by the following variables: a \emph{starting point}, a \emph{surface direction}, and a \emph{length and width}. The \emph{starting point} is defined using barycentric coordinates $(u_e, v_e)$ on a particular element $e$ of the garment mesh. The \emph{surface direction} is a vector in barycentric space $(\overrightarrow{u_e}, \overrightarrow{v_e})$ from the starting point to another barycentric coordinate on the same element $e$.

We start by tracing out a piece-wise linear path of the desired length in the direction of $(\overrightarrow{u_e}, \overrightarrow{v_e})$ until an edge is encountered, whereby the vector is converted to Euclidean space $\in\mathbb{R}^{3n}$ and rotated to lie on  the surface of the next triangle $e_i$. This is repeated until the length of the vector is exhausted. Two endpoints are produced, one at the starting point, and one at the last barycentric coordinate of where the path finishes.

From this path, a mesh is triangulated by creating center vertices at edge intersections and projecting side vertices to the left and right of the path based on $\overrightarrow{e_n} \times \overrightarrow{p_{xyz}}$, the cross product of the element normal and the path direction in world space respectively. This is scaled by the $width$ parameter.  The resulting mesh has rest vertices $\mathbf{q}=(\mathbf{q}_1,\ldots,\mathbf{q}_m)\in\mathbb{R}^{3n}$ and $\bar{\mathbf{q}}=(\bar{\mathbf{q}}_1,\ldots,\bar{\mathbf{q}}_m)\in\mathbb{R}^{3n}$ when deformed. 
We give special treatment to the side vertices of the two endpoints by walking them in an orthogonal direction to the main path using the same walking algorithm outlined above. 

The endpoints $q^c$ of the clutch mesh (3 at each end) are connected to the garment mesh using simple quadratic penalty functions, which allows for firm attachment. The full path walking and meshing algorithm is fast enough to work in real-time, allowing for rapid user placement and re-positioning of ES clutches.  


{\vfill\pagebreak}
\section{Garment-on-Body Model} 
\label{ap:sim_model}

As our garment model, we use a compressible neo-Hookean material model \cite{bonet_wood_2008} adapted with a relaxed energy under wrinkling as in \cite{vechev2022cdkg}. This allows the garment to wrinkle under compression without producing geometric artifacts. This results in the garment energy $E_\mathrm{garment}(\bx, \materialdensity)$, which is a function of the garment design $\materialdensity$, and the deformed nodal positions $\bx$. We similarly convert the discrete body mesh to a continuous implicit signed distance field \cite{oztireli2009feature}, resulting in the energy $E_\mathrm{body}(v, \bx)$, which pushes back on the garment vertices $\bx$ away from the body. This allows the garment to smoothly slide on top of the body and to lift-off from its surface. To attach the garment to the body in specific areas, we introduce a simple coupling potential, $E_\mathrm{attach} = \frac{1}{2}k(x^c-x^v_c)^T(x^c-x^v_c)$, attracting elements of the garment mesh $x^c$ to corresponding elements $v_c$ on the body mesh. As the garment mesh is initialized from the SMPL mesh, for more accurate simulation, we subdivide the garment mesh until it has 16x the resolution of the base SMPL template mesh.


\section{On-Body Topology Optimization} 
\label{ap:onbody_tpo}

To design passive reinforcement structures, Vechev et al. use a bi-directional evolutionary structural optimization (BESO) algorithm \cite{huang2007convergent, huang2009bi} to solve the constrained optimization problem with a single objective, 

\begin{equation}
\label{eq:BESOObjective}
\begin{split}
\bd^* = \arg\max_{\bd}  \quad E_\mathrm{garment}(x^*, \bd) \\ 
\textrm{s.t.} \quad \sum_e{A^e} d^e=A^*\ ,\quad \bff(x^*)= \mathbf{0} .
\end{split}
\end{equation}
The goal of this formulation is to find an optimal per-element material assignment $\bd^*$ that maximizes the energy of the garment in its equilibrium state $\bx^*$ while satisfying constraints on force equilibrium, $\bff(x^*)= \mathbf{0}$, and material budget, $\sum_e{A_e} d_e=A^*$. The strain energy of the garment is defined per element as
\begin{equation}
\label{eq:GarmentEnergy}
E_{\mathrm{garment}} = \sum_e t^e  A^e\energydensity_\mathrm{garment}^e(x^*, \materialdensity^e)\ , 
\end{equation}
where $\energydensity_\mathrm{garment}^e$ is the elemental strain energy density, and $t^e, A^e$ are the thickness and area of the element respectively.

\end{document}